# A Simple Numerical Absorbing Layer Method in Elastodynamics

*Jean-François Semblat, Ali Gandomzadeh, Luca Lenti*

*Université Paris-Est, LCPC, 58 bd Lefebvre, 75015 Paris, France*

**Abstract:** The numerical analysis of elastic wave propagation in unbounded media may be difficult to handle due to spurious waves reflected at the model artificial boundaries. Several sophisticated techniques such as nonreflecting boundary conditions, infinite elements or absorbing layers (e.g. Perfectly Matched Layers) lead to an important reduction of such spurious reflections. In this Note, a simple and efficient absorbing layer method is proposed in the framework of the Finite Element Method. This method considers Rayleigh/Caughey damping in the absorbing layer and its principle is presented first. The efficiency of the method is then shown through 1D Finite Element simulations considering homogeneous and heterogeneous damping in the absorbing layer. 2D models are considered afterwards to assess the efficiency of the absorbing layer method for various wave types (surface waves, body waves) and incidences (normal to grazing). The method is shown to be efficient for different types of elastic waves and may thus be used for various elastodynamic problems in unbounded domains.

**Abstract:** La simulation numérique de la propagation d'ondes élastiques en milieux infinis peut s'avérer délicate du fait des réflexions d'ondes parasites sur les frontières du modèle discrétisé. Plusieurs méthodes sophistiquées, telles que les frontières absorbantes, les éléments infinis ou les couches absorbantes (e.g. « Perfectly Matched Layers ») permettent une réduction importante des réflexions parasites. Dans cette note, une méthode de couche absorbante simple et efficace est proposée dans le cadre de la méthode des éléments finis. Cette méthode s'appuie sur une formulation de l'amortissement dans la couche de type Rayleigh/Caughey et ses principes sont d'abord détaillés. L'efficacité de la méthode est alors démontrée grâce à des simulations unidimensionnelles en considérant un amortissement homogène ou variable dans la couche absorbante. Des modèles bidimensionnels permettent ensuite d'apprécier l'efficacité de la méthode de couche absorbante proposée pour différents types d'ondes (ondes de surface, ondes de volume) et des incidences variées (normale à rasante). La méthode s'avère ainsi efficace pour différents types d'ondes élastiques et pourrait être utilisée pour traiter divers problèmes élastodynamiques en milieux non bornés.

## 1. Numerical methods for elastic waves in unbounded domains

Various numerical methods are available to simulate elastic wave propagation in solids: finite differences [1,2], finite elements [3,4], boundary elements [5,6,7], spectral elements [8,9], etc. Such methods as finite or spectral elements have strong advantages (for complex geometries [9], nonlinear media [10], etc) but may have such drawbacks as numerical dispersion [4,11,12] for low order finite elements, or spurious reflections at the mesh boundaries [4,13].

The problem of spurious reflections may be dealt with using the Boundary Element Method [5,6,7] or coupling it with other numerical methods [14]. Domain Reduction Methods are also available in the framework of Finite Element approaches [15] but are mainly dedicated to large scale simulations.

Another alternative is to directly attenuate the spurious reflections at the mesh boundaries considering appropriate conditions at the boundary or in a fictitious absorbing medium. Absorbing boundaries techniques, or nonreflecting boundary conditions (NRBC), involve specific boundary conditions to approximate the radiation condition for the elastic waves [4,13,16,17,18,19]. Such techniques generally involve spatial and temporal derivatives of the dynamic motion and may be efficient for both the harmonic and the time-dependent cases [18,19]. The infinite elements formulation directly allows the approximation of the decaying laws governing the radiation process at infinity [4,20]. Absorbing layers methods, such as Perfectly Matched Layers (*PML*) [4,21,22,23,24,25] are based on the description of attenuating properties along a specific direction in a fictitious absorbing layer of finite thickness located at the medium boundaries. Classical PMLs are generally efficient but several cases lead to severe instabilities: grazing incidence, shallow models involving surface waves, anisotropic media [25]. The multi-directional PML formulation [4,25] allows various choices for the attenuation vector $\underline{\alpha}$ in the absorbing layer:

$$\underline{u} = \underline{A} \exp(-\underline{\alpha}.\underline{x}) \exp[-i(\underline{k}.\underline{x} - \omega t)] \qquad (1)$$

where $\underline{u}$ is the displacement vector, $\underline{x}$ the position vector, $\underline{k}$ the wave vector, $\omega$ the circular frequency, $t$ the time.

It is thus possible to deal with various wave types (i.e. polarizations) and incidences. The multi-directional PML formulation leads to better efficiency and numerical stability [4,25]. However, since the attenuating properties are described by an analytical law, the implementation of the PML technique may be difficult and its numerical cost may also be significant.

In this Note, we propose a simple and efficient absorbing layer method to reduce the spurious wave reflections in unbounded elastodynamics. Beyond its simplicity, the interest of the method is that it is independent of the wave incidence and already available in most of the general purpose Finite Element softwares (no specific implementation required).

## 2. A Simple Multi-Directional Absorbing Layer Method

### 2.1 Discretized equation of motion

Considering linear elastodynamics (small strains) and using variational principles, the equation of motion is discretized in space as follows [3,4]:

$$[M]\{\ddot{u}\} + [C]\{\dot{u}\} + [K]\{u\} = \{0\} \qquad (2)$$

with the following variable boundary condition at $\underline{x}_i$ ($i=j,\ldots,k$ for $k-j+1$ nodes where the condition is defined):

$$u_\ell^{(i)}(t) = R_2(t, t_s, t_p) \qquad (3)$$

[$M$], [$C$], [$K$] being the mass, damping and stiffness matrices respectively, $\{u\}$ the vector of nodal displacements, $u_\ell^{(i)}(t)$ the $\ell^{\text{th}}$ component of displacement at $\underline{x}_i$, $R_2(t,t_s,t_p)$ a Ricker wavelet [4] and $t_s, t_p$ its parameters (time shift and fundamental period respectively).

Isoparametric finite elements are considered herein. The time integration scheme is a Newmark non dissipative one to avoid algorithmic damping [3].

### 2.2 Basic idea of the absorbing layer method

Since the better ideas to deal with spurious reflections are to consider absorbing layers and multi-directional attenuating properties (Eq. (1)), our goal is to proceed as for multi-directional PML formulation [25] but with a simple description of the attenuation process. Physical models are generally very efficient to describe the attenuation process (frequency dependence, causality, etc.) but they are nevertheless not very easy to implement or cost effective due to the use of memory variables for instance [10,26,27,28,29]. We propose to use the Caughey damping formulation [3,4] to describe the attenuation of the waves in an absorbing layer of finite thickness. This damping formulation is purely numerical but, as recalled hereafter, a rheological interpretation may be proposed (i.e. 2$^{\text{nd}}$ order Caughey formulation).

### 2.3 Rayleigh and Caughey damping

Rayleigh damping is a classical method to easily build the damping matrix [$C$] for a Finite Element model [3,4] under the following form:

$$[C] = a_0[M] + a_1[K] \qquad (4)$$

where [$M$] and [$K$] are the mass and stiffness matrices of the whole model respectively. It is then called *Rayleigh damping matrix* and $a_0, a_1$ are the Rayleigh coefficients. [$C$] is the sum of two terms: one is proportional to the mass matrix, the other to the stiffness matrix.

A more general damping formulation was proposed by Caughey [30,31] as follows:

$$[C] = [M]\sum_{j=0}^{m-1} a_j \left([M]^{-1}[K]\right)^j \qquad (5)$$

As evidenced by Eq. (5), the Rayleigh formulation (Eq. (4)) corresponds to a 2$^{\text{nd}}$ order Caughey damping ($m=2$) involving a linear combination of the mass and stiffness matrices.

These formulations of the damping matrix are very convenient since it can be easily computed. Furthermore, for modal approaches, the *Rayleigh (or Caughey) damping matrix* is diagonal in the real modes base and uncoupled damped modal equations may thus be derived [4,32].

Rayleigh (or Caughey) damping formulation may also be used to analyze the propagation of damped elastic wave [4]. For such problems, it may be useful to have a rheological interpretation of these purely numerical formulations. In the field of mechanical wave propagation, the equivalence between the Rayleigh formulation and a Generalized Maxwell model, proposed in [33], is now briefly recalled.

### 2.4 Rheological interpretation of Rayleigh damping

Considering Rayleigh damping [3] (i.e. 2$^{nd}$ order Caughey damping), the damping ratio $\xi$ may be written as follows:

$$\xi = \frac{a_0}{2\omega} + \frac{a_1 \omega}{2} \tag{6}$$

where $\omega$ is the circular frequency and $a_0$, $a_1$ are the constant Rayleigh coefficients.

The equivalence between Rayleigh damping and the damping properties of a Generalized Maxwell fluid (Fig. 1) is evidenced in [33] for wave propagation problems (through the complex modulus and quality factor of the latter). The Rayleigh coefficients $a_0$ and $a_1$ may thus be identified from experimental results through the rheological parameters. This equivalence only characterizes the damping properties (damping matrix) in the Finite Element model; the elastic behaviour is mainly controlled by the stiffness matrix.

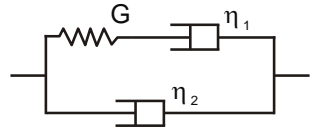

**Fig. 1.** Rheological model for a Generalized Maxwell fluid leading to the same damping properties as the Rayleigh formulation [33,34].

### 2.5 Proposed absorbing layer method

In the framework of the Finite Element Method, an elastic medium is considered and an absorbing layer system will be designed at its boundaries. The absorbing layer is thus modelled with appropriate damping properties (i.e. Rayleigh/Caughey damping coefficients) in order to attenuate the spurious reflections at the mesh boundaries. This simple absorbing layer method may thus reduce the amplitude of the elastic waves coming from the elastic medium and reflecting at the artificial boundaries of the domain. In the following, the proposed technique is described for different damping variations in the absorbing layer thickness. The spatial variations of damping are controlled by variable damping coefficients in the Finite Elements. Such techniques were already used to model wave propagation in media with stress state dependent damping [36,37]. Considering Rayleigh damping, the element damping matrix for finite element "$e$" is thus written:

$$[C^{(e)}] = a_0^{(e)}[M^{(e)}] + a_1^{(e)}[K^{(e)}] \tag{7}$$

The Rayleigh damping coefficients may be different in each finite element or chosen piecewise constant in the absorbing layer. In the following, the efficiency of the proposed absorbing layer method is assessed for 1D and 2D elastic wave propagation.

## 3. Efficiency of the 1D absorbing layer

### 3.1 Definition of the propagating wave

For the 1D case, a 2$^{nd}$ order Ricker wavelet is considered [4,38]. It is derived from a Gaussian and is well localized in both time and frequency domain. We shall consider longitudinal elastic waves and the predominant frequency of the Ricker wavelet $f_R = 1/t_p$ will be chosen in order to have an integer number of wavelengths along the medium.

### 3.2 Rayleigh damping in the absorbing layer

In the following, the absorbing layer involves Rayleigh damping (2$^{nd}$ order Caughey damping) which is frequency dependent and is rheologically equivalent to a Generalized Maxwell fluid (Fig. 1) for wave propagation problems [33]. From Eq. (6), a band-pass response is evidenced for the Rayleigh damping (minimum value at a given frequency). To define a reference attenuation (inverse of the quality factor: $Q^{-1}$) in the absorbing layer, the minimum attenuation will be chosen at the predominant frequency of the Ricker wavelet $f_R$. In the present examples, the efficiency of the absorbing layer will thus be the lowest at the chosen predominant frequency (lowest attenuation). Due to the band-pass behaviour, the spurious waves will be more attenuated at frequencies lower and higher than $f_R$.

From the expression of the damping ratio, Eq. (6), it is possible to determine the frequency of minimum damping $\omega_{min}$ from the Rayleigh coefficients as follows:

$$\omega_{min} = \sqrt{\frac{a_0}{a_1}} \quad (8)$$

Choosing the minimum damping $\xi_{min}$ (or attenuation $Q_{min}^{-1} \approx 2\xi_{min}$ [33,35]) at the predominant frequency of the Ricker wavelet $f_R$, it is then possible to derive the following relation:

$$\omega_R = 2\pi f_R = \sqrt{\frac{a_0}{a_1}} \quad (9)$$

and thus, using the definition of Rayleigh damping, derive the following system:

$$\begin{cases} Q_{min}^{-1} = 2\xi_{min} = \dfrac{a_0}{\omega_R} + a_1 \omega_R \\ \omega_R = 2\pi f_R = \sqrt{\dfrac{a_0}{a_1}} \end{cases} \quad (10)$$

The choice of the predominant frequency of the Ricker wavelet $f_R$ and the minimum attenuation $Q_{min}^{-1}$ thus allows the estimation the Rayleigh damping coefficients in the absorbing layer. In the following, we shall choose several typical values for $Q_{min}^{-1}$ ranging from 0.5 (i.e. $\xi_{min}=0.25$) to 1.0 (i.e. $\xi_{min}=0.5$).

### *3.3 Homogeneously absorbing case*

#### 3.3.1 Description of the homogeneous absorbing layer system

As depicted in Fig. 2, the first numerical case corresponds to a 1D elastic medium and a homogeneously absorbing layer (shaded). The length of the elastic layer (left) is $4\lambda$ and that of the homogeneously damped layer (right) is $\lambda$ ($\lambda=c/f_R$ wavelength of the longitudinal wave and $c$ wave velocity). The size of the finite elements is $\lambda/20$ to have low numerical dispersion [11,12]. In the elastic medium, the element damping matrices are zero whereas homogeneous Rayleigh damping is considered in the absorbing layer by choosing identical Rayleigh coefficients for each element damping matrix in this area (the elastic properties being identical in both domains). In each case, the attenuation value $Q^{-1}$ is chosen as the minimum attenuation value at the predominant frequency of the propagating wave (Ricker wavelet). Two cases are considered: $Q_{min}^{-1}=0.5$ and 1.0.

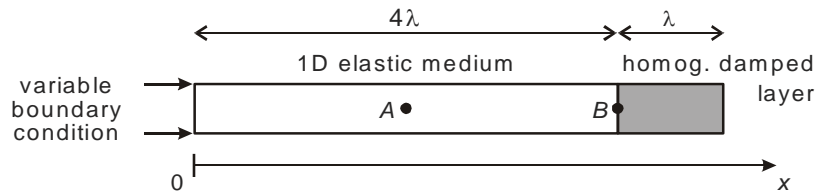

**Fig. 2.** Schematic of the first numerical test: undamped elastic layer (left) and homogenously damped layer (right)

#### 3.3.2 Efficiency of the homogeneously damped layer

For the two different attenuation values in the absorbing layer, the results are displayed in Fig. 3 for point *A* (left) at the centre of the elastic medium and point *B* (right) at the interface between the elastic medium and the absorbing layer.

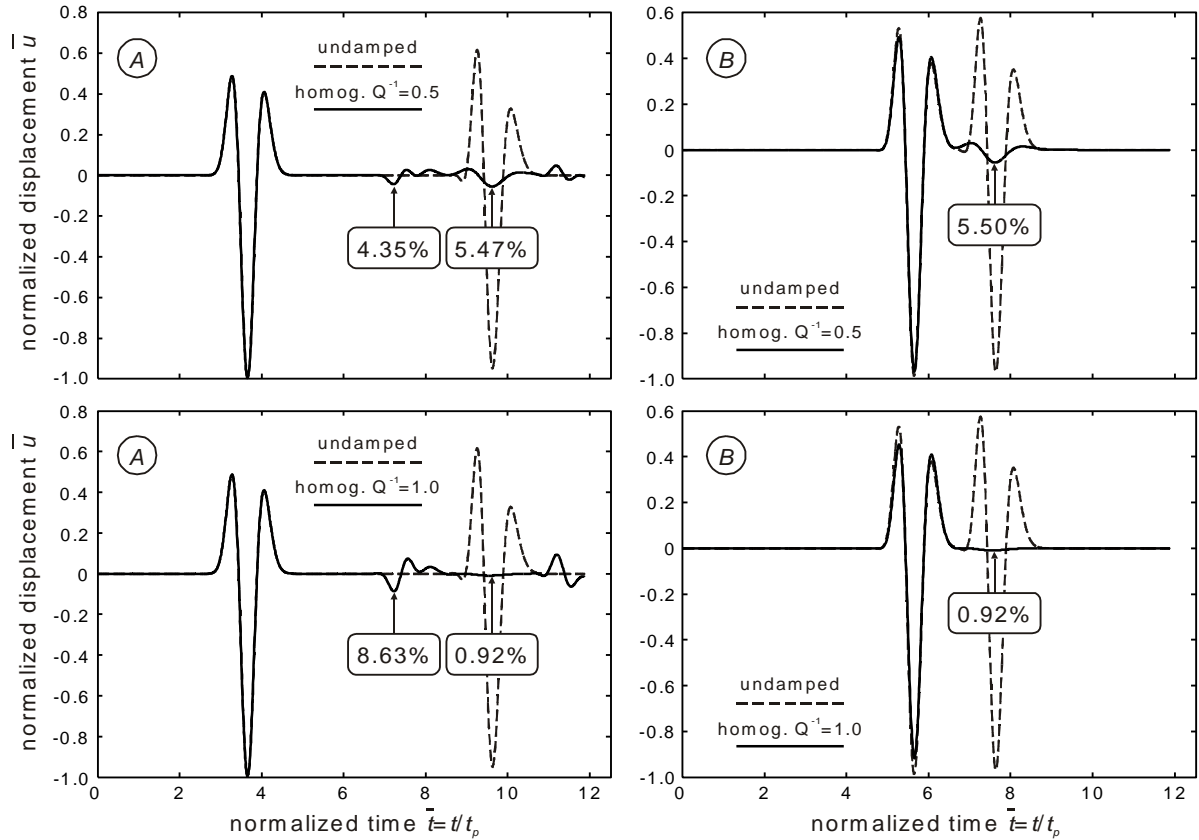

**Fig. 3.** Comparison between the homogeneously damped layer (solid) and the undamped case (dashed) at points *A* (left) and *B* (right) for different attenuations: $Q_{min}^{-1}$ =0.5 (top) and $Q_{min}^{-1}$ =1.0 (bottom).

These curves are plotted in terms of normalized displacement $\bar{u} = u_x / A_x$ vs normalized time $\bar{t} = t / t_p$ and lead to the following conclusions ($u_x$ is the displacement along *x* and $A_x$ is the maximum amplitude of the incident wave):

- For $Q_{min}^{-1}$ =0.5 (top, solid), when compared to the undamped case ($Q_{min}^{-1}$ =0.0, dashed), the amplitude of the reflected wave at point *A* is much smaller (5.47% of $A_x$) but the incident wave is also reflected at the interface between the elastic medium and the absorbing layer (4.35%). It is due to the contrast between the elastic medium and the viscoelastic layer in terms of complex modulus or complex wavenumber [4,35]. For $Q_{min}^{-1}$ =0.5 at point *B* (top right, solid), the amplitude at the interface between the elastic medium and the absorbing layer is also very small (5.50%). The efficiency of the homogeneously absorbing layer thus appears satisfactory.

- For $Q_{min}^{-1}$ =1.0 (bottom, solid), the amplitude of the reflected wave at the end of the absorbing layer is nearly zero (0.92% of $A_x$) but the reflected wave at the interface with the elastic medium is larger than for $Q_{min}^{-1}$ =0.5 (8.63%). This is due to the fact that the complex velocity contrast is larger for $Q_{min}^{-1}$ =1.0.

Similar results were obtained for transverse waves but are not reproduced here. From these homogeneously damped cases, the proposed absorbing layer method can already be considered as an efficient method but its efficiency may probably be improved and its artefacts reduced.

### *3.4 Continuously varying damping*

3.4.1 Description of the continuous absorbing layer system

The second numerical case corresponds to a continuously varying damping in the absorbing layer. The absorbing layer involves variable Rayleigh damping coefficients increasing linearly with the horizontal distance. The idea is to have a continuously increasing damping value in the absorbing layer system and to match the (zero) damping values at the interface with the elastic medium. The attenuation $Q_{min}^{-1}$ thus increases linearly from 0 at $x=4\lambda$ to 1.0 at $x=5\lambda$.

3.4.2 Efficiency of the continuously absorbing layer system

For the continuously (i.e. linearly) damped case, the results at point *A* (centre of the elastic medium) are displayed in Fig. 4 and compared to the undamped case ($Q_{min}^{-1}$=0.0). The efficiency of the continuous absorbing layer system is much better than that of the homogeneous absorbing layer for the reflection at the interface: 1.11% of the maximum amplitude of the incident wave instead of 8.63%. For the reflection at the medium boundary, the efficiency of the continuous system is a bit less: 5.25% instead of 0.92%. However, when compared to the homogeneous case, the overall efficiency of the continuous system is very good.

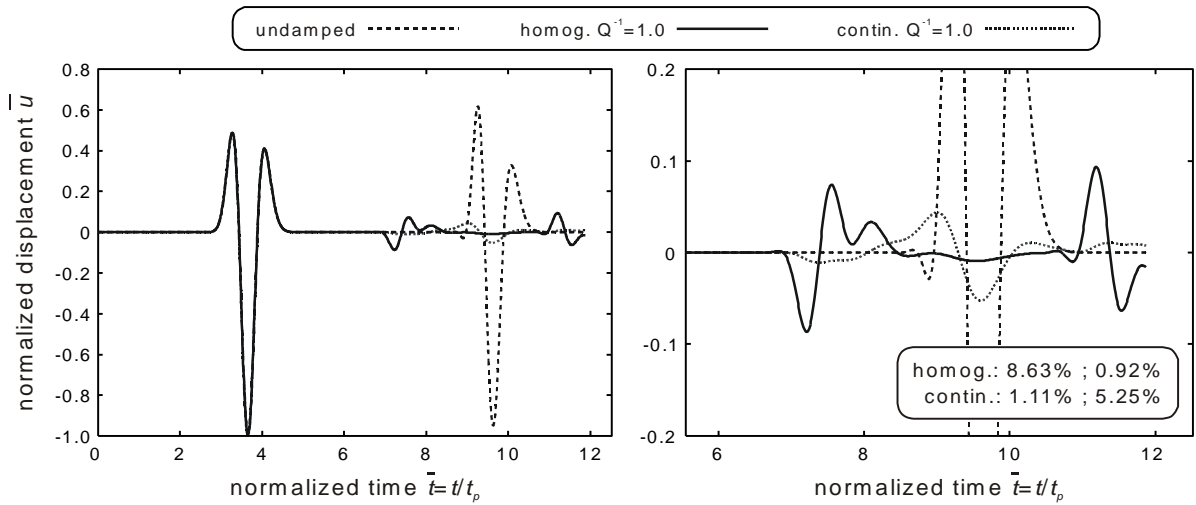

**Fig. 4.** Comparison between the continuous absorbing layer, the homogeneous absorbing layer and the undamped case at point *A*. Closer view at right.

## 4. Efficiency of the 2D absorbing layer

### *4.1 Definition of the propagating wave and geometry*

To assess the efficiency of the proposed absorbing layer method in 2D, a 2D FEM model is considered (Fig. 5). A plane strain model is chosen in order to avoid strong geometrical damping. It involves a $4\lambda \times 4\lambda$ square elastic medium and two $\lambda$ thick absorbing layers (right and bottom). The model is symmetrical along the left boundary and the variable boundary condition (2nd order Ricker wavelet at $f_R$ [4]) is applied along a distance of $\lambda/2$ at the free surface. The wavefield in the model is thus composed of various wavetypes (longitudinal, transverse and surface waves) and the motion duration is larger than in the 1D case. The element size, $\lambda/20$, is identical to that of the 1D case.

The efficiency of the homogeneous and continuous damped cases will be assessed in the whole model (isovalues plots at some selected times) and at point *A* along the free surface (normalized displacement vs normalized time).

In the work of Festa and Vilotte [23], various 2D cases are considered for PML formulations (pure body waves, thin layer, curved layer).

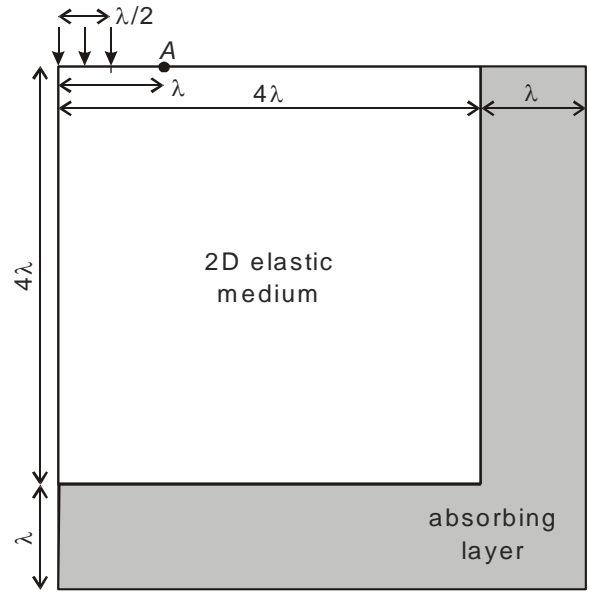

**Fig. 5.** Schematic of the 2D plane strain model showing the boundary condition (variable vertical displacement) and the 2D absorbing layer.

### 4.2 Efficiency of the homogeneous absorbing layers

The results obtained in the homogeneous case are displayed in Fig. 6 for $Q_{min}^{-1}$ =1.0 in terms of normalized displacement (norm of the displacement vector divided by the maximum amplitude of the Ricker wavelet) at three different normalized times ($\bar{t}_1 = 4.15$, $\bar{t}_2 = 9.135$ and $\bar{t}_3 = 14.1$). For this displacement isovalue scale ($0.0 \leq \bar{u} \leq 0.6$), the reflected waves are obvious in the undamped case (top). For the homogeneously damped case (bottom), when compared to the amplitudes of the undamped case, the first reflections are not significant at time $\bar{t}_2$ (only the incident wavefield is present in this plot) and no reflection at all can be identified at time $\bar{t}_3$.

Since the displacement isovalues scale is rather large in Fig. 6, a quantitative analysis is now proposed in terms of norm of the displacement vector divided by the maximum amplitude in the undamped case at the same time. The amplitude ratio of the reflected wave was also computed for $Q_{min}^{-1}$ =0.5: it reaches 2.48% of the maximum amplitude obtained in the undamped case. For $Q_{min}^{-1}$ =1.0, the homogeneously absorbing layer system leads to larger amplitudes of the reflected waves (4.67%) surely due to the velocity contrast at the interface (see results from the 1D homogeneous case).

The efficiency of the proposed absorbing layer method thus appears satisfactory in the 2D homogeneous case but the reflections at the interface should be reduced.

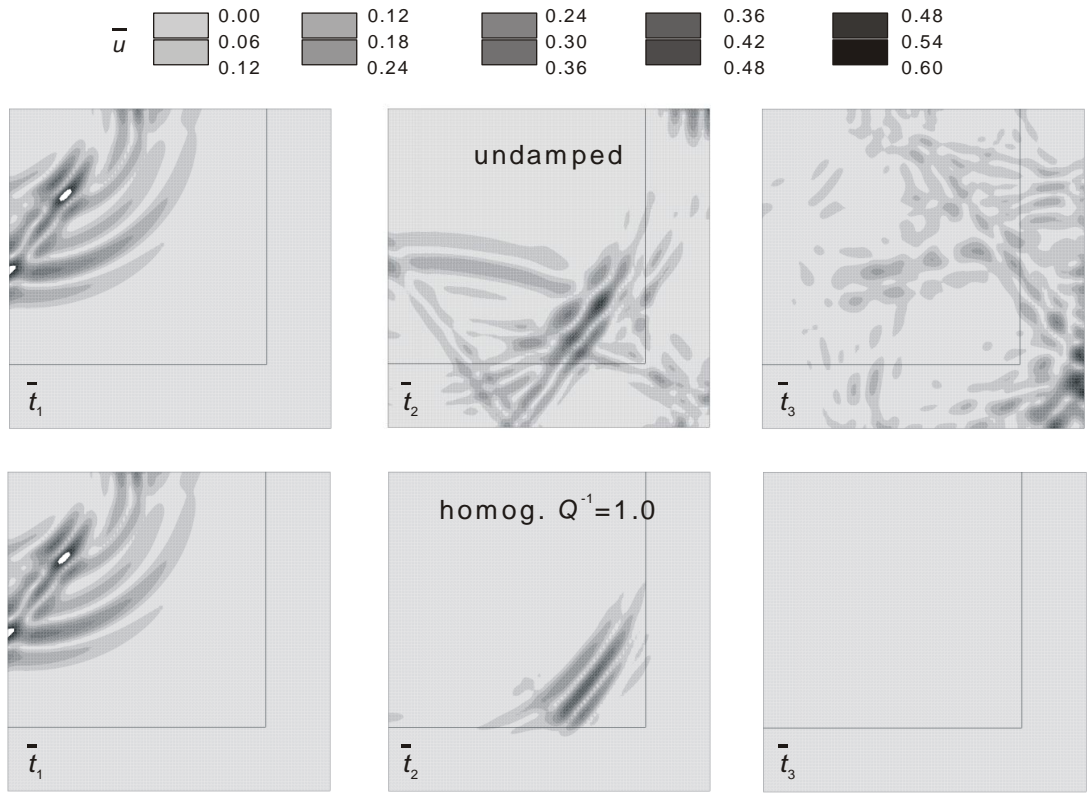

**Fig. 6.** Comparison between the undamped case (top) and the homogeneous case ($Q_{min}^{-1}$ =1.0, bottom) at three different times: normalized displacement at $\bar{t}_1 = t_1 / t_p = 4.15$ (left) ; $\bar{t}_2 = t_2 / t_p = 9.13$ (centre) ; $\bar{t}_3 = t_3 / t_p = 14.1$ (right).

### 4.3 *Efficiency of the continuous absorbing layers*

To improve the efficiency of the proposed 2D absorbing layer method, the continuous absorbing case is now considered. The numerical results at the free surface (point *A*, Fig. 5) are now displayed in terms of normalized displacement vs normalized time (Fig. 7). All 2D cases are considered: homogeneous case for $Q_{min}^{-1}$ =0.5 and $Q_{min}^{-1}$ =1.0 (top) and continuous case for $Q_{min}^{-1}$ =1.0 (bottom). As in the 1D case, the reflections at the interface and at the medium boundaries are quantified in terms of maximum relative amplitude.

In the homogeneous case (Fig. 7, top), the surface waves reflected at the interface have large amplitudes when compared to the undamped case (6.04% for $Q_{min}^{-1}$ =0.5 to 12.0% for $Q_{min}^{-1}$ =1.0). The reflected waves at the medium right boundary reach lower amplitudes (5.60% and 7.18% resp.).

In the continuous case for $Q_{min}^{-1}$ =1.0 (Fig. 7, bottom), the reflection at the interface leads to low amplitudes (1.21%) whereas the reflections at the medium boundaries are a bit larger (5.01%) but nevertheless lower than in the homogeneous case. As shown by the time histories, the variations with time are strong and it is difficult to quantify a maximum amplitude in both time and space.

When compared to the PML method proposed by Festa and Vilotte [23], the efficiency of the present method may appear lower. However, the latter is independent of the wave incidence. Indeed, for grazing incidences, the standard PML technique leads to much lower amplitude reductions and the present method may then be more efficient. In such cases, an alternative to standard PML techniques is to consider a multi-directional PML formulation [25] to have a free choice of the attenuation vector in the absorbing layer. However, the present method is very simple and it is intrinsically multidirectional.

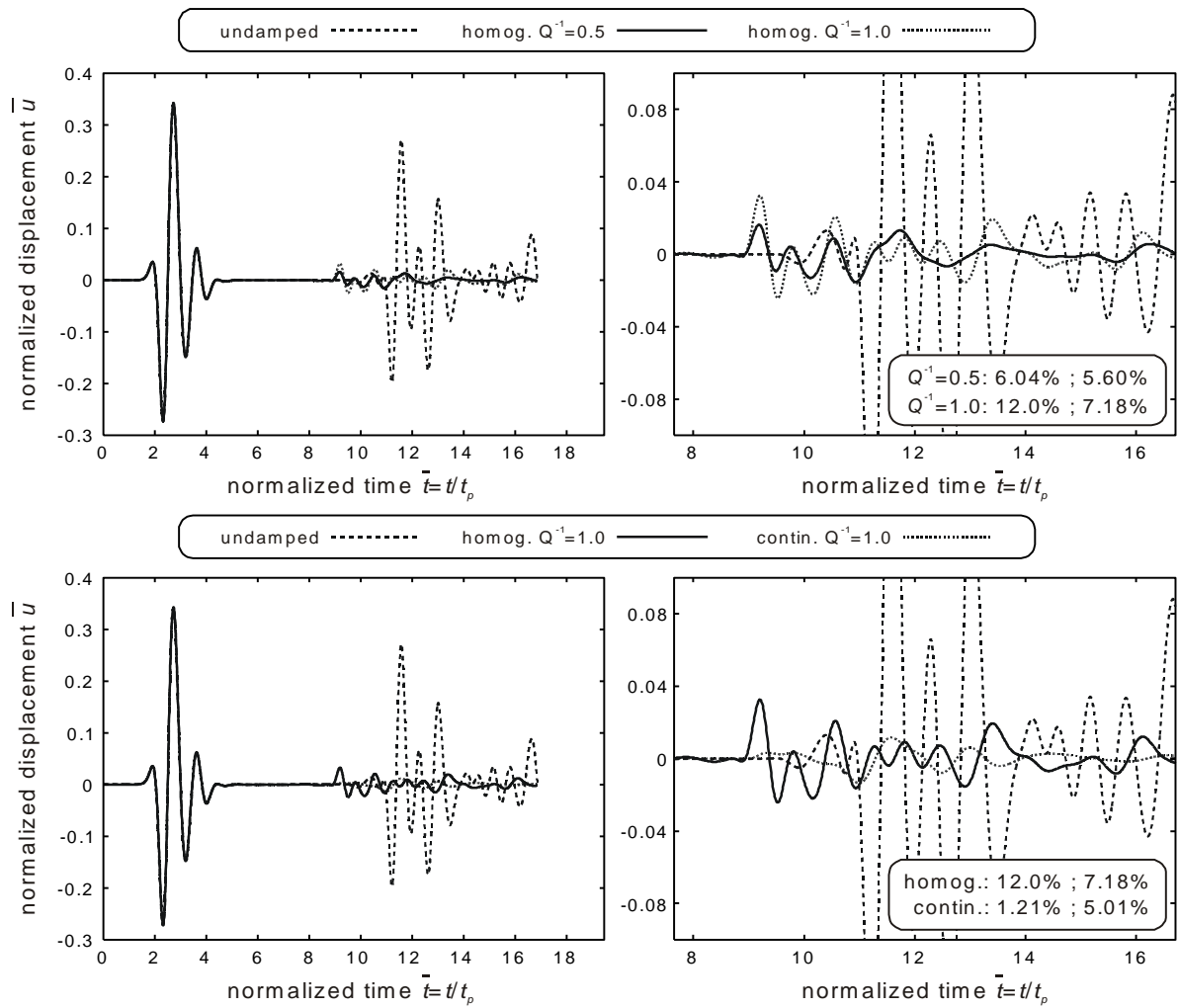

**Fig. 7.** Comparison between homogeneous, heterogeneous and continuous cases at point *A* (surface, Fig. 5) and maximum amplitude ratio of the reflected waves.

## 5. Conclusion

The main conclusion of this work is that the proposed absorbing layer method is efficient to reduce the spurious wave reflections in Finite Element models for unbounded elastodynamics. Furthermore, when compared to other absorbing boundary/layer techniques or viscoelastic mechanical models, it is independent of the wave incidence and is generally available in most of the general purpose Finite Element softwares.

In future works, it will be necessary to assess the efficiency of the proposed method in 3D realistic cases. It may be also useful to consider higher order Caughey damping (Eq. (5)) leading to various types of damping-frequency variations. In addition to the Finite Element Method, the method may be considered in the framework of other numerical methods such as the Spectral Element Method [8,9].